\documentclass[12pt]{article}%
\usepackage{citesort}
\usepackage{amsmath}
\usepackage{graphicx}%
\usepackage{amsfonts}%
\usepackage{amssymb}
\begin{document}

\title{Forward-backward multiplicity correlations \\in the wounded nucleon model}
\author{Adam Bzdak\thanks{e-mail: Adam.Bzdak@ifj.edu.pl}\\Institute of Nuclear Physics, Polish Academy of Sciences\\Radzikowskiego 152, 31-342 Krakow, Poland}
\maketitle

\begin{abstract}
The forward-backward multiplicity correlation strength is calculated for
arbitrary nucleus-nucleus collision in the framework of the wounded nucleon
model. Discussion of our results in the context of the recent STAR data in
$AuAu$ collisions at $\sqrt{s}=200$ GeV is presented. It is suggested that the
observed (i) growth of the correlation coefficient with centrality and (ii)
approximately flat pseudorapidity dependence of the correlation strength for
central collisions are due to the fluctuations of the number of wounded
nucleons at a given centrality bin.

\vskip 0.6cm


\end{abstract}

\section{Introduction}

Recently the STAR collaboration announced the results \cite{STAR} on the
forward-backward multiplicity correlations in nucleus-nucleus collisions. It
was found that the correlation strength (defined below) was larger than in an
elementary proton-proton collisions and it remains constant (at least for the
most central collisions) across the measured midrapidity region. This result
was interpreted in the framework of the color glass condensate \cite{CGC} or
dual parton \cite{DPM} models, which suggests the possible formation of high
density partonic matter in central $AuAu$ collisions at $\sqrt{s}=200$ GeV.
Other theoretical investigations concerning the problem of forward-backward
multiplicity correlations in hadronic collisions can be found in Refs.
\cite{Torr,BPV,BDP-lhc,GU,Chou,Lim,FK,BBP,Yan}.

The main difficulty, however, is to distinguish between correlations arising
from the presence of the quark-gluon plasma and correlations that do not
depend on this new phenomenon. These need to be understood, controlled and
subtracted in order to access the true signal of the quark-gluon plasma. The
natural ground to study this problem is the wounded nucleon model \cite{WNM}.
Indeed, it is the simplest superposition model in which a nucleus-nucleus
collision is constructed from an elementary nucleon-nucleon collisions. More
precisely, the number of produced particles in nucleus-nucleus collision is
proportional to the number of wounded nucleons, i.e., nucleons that underwent
at least one inelastic collision. A Monte Carlo analysis of this problem in
the very simplified wounded nucleon model was already presented in Ref.
\cite{Torr}.

Our main conclusion is that the STAR data \cite{STAR} can be naturally
understood in the wounded nucleon model (at least for the most central
collisions). We conclude that the observed growth of the correlation
coefficient with centrality and approximately flat pseudorapidity dependence
of the correlation strength are due to the fluctuations of the number of
wounded nucleons at a given centrality bin.

\begin{figure}[h]
\begin{center}
\includegraphics[scale=1.5]{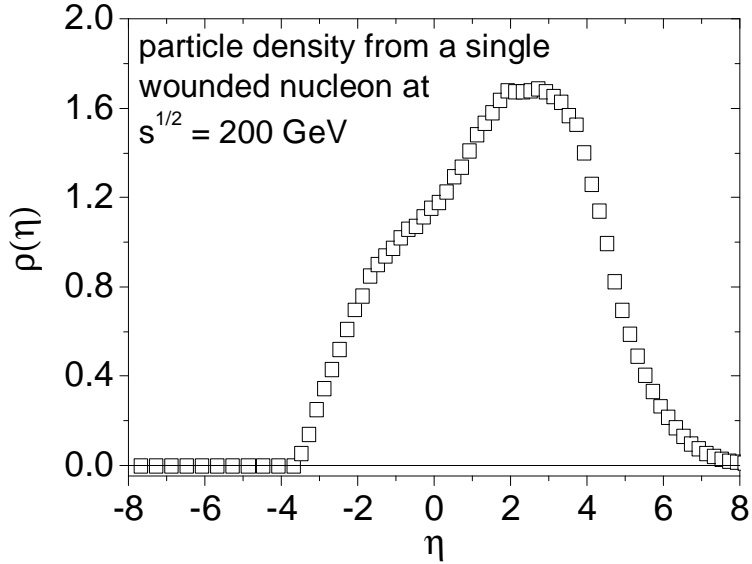}
\end{center}
\caption{A wounded nucleon fragmentation function at $\sqrt{s}=200$ GeV c.m.
energy.}%
\label{fig_n}%
\end{figure}

The correlation coefficient (or correlation strength) $b$ is defined as%
\begin{equation}
b=\frac{\left\langle n_{B}n_{F}\right\rangle -\left\langle n_{B}\right\rangle
\left\langle n_{F}\right\rangle }{\left\langle n_{F}^{2}\right\rangle
-\left\langle n_{F}\right\rangle ^{2}}\equiv\frac{U}{D}, \label{b_def}%
\end{equation}
where $n_{B}$ and $n_{F}$ are event by event particle multiplicities in
backward $B$ and forward $F$ pseudorapidity\footnote{Our discussion is valid
for any longitudinal variable, not necessarily pseudorapidity.} intervals,
respectively. The main ingredients which allow to evaluate $b$ in the wounded
nucleon model are (i) recently obtained pseudorapidity particle density from a
wounded nucleon $\rho(\eta)$ and (ii) particle multiplicity distributions
measured in $pp$ collisions in different forward and backward intervals. The
fragmentation function $\rho(\eta)$, shown in Fig. \ref{fig_n}, was obtained
by analysing the PHOBOS data on $dAu$ collisions \cite{PHO-dAu} at $\sqrt
{s}=200$ GeV in the wounded nucleon model \cite{ff-bc} and the wounded
quark-diquark model\footnote{In this case $\rho\left(  \eta\right)
=1.2F(\eta)+0.8U(\eta)$, where $F(\eta)$ and $U(\eta)$ are the particle
densities from wounded and unwounded constituents, respectively.}
\cite{ff-bb}. In a completely independent way, mainly based on the recent NA49
collaboration data \cite{NA49}, analogous wounded nucleon fragmentation
function was constructed in Ref. \cite{ff-Ryb}. For the multiplicity
distributions measured in $pp$ collisions we take the negative binomial (NB)
fits \cite{UA5-200-n-k}%
\begin{equation}
P(n,\bar{n},k)=\frac{\Gamma(n+k)}{\Gamma(n+1)\Gamma(k)}\left(  \frac{\bar{n}%
}{k}\right)  ^{n}\left(  1+\frac{\bar{n}}{k}\right)  ^{-n-k}, \label{nbd}%
\end{equation}
where $\bar{n}$ is the average multiplicity and $1/k$ measures deviation from
Poisson distribution.

In the next section the correlation coefficient $b$ for the collision of two
arbitrary nuclei is derived. In section $3$ we focus on the collision of two
symmetric nuclei and look closer at the midrapidity and fragmentation regions,
where $b$ can be written in a particularly simple form. Our results are
discussed in the context of the recent STAR data in section $4$ and section
$5$ where also some comments are included. In the last section our conclusions
are listed.

\section{Model}

The problem is to calculate the correlation coefficient (\ref{b_def}) in two
given pseudorapidity regions $B$ and $F$ under assumption that the
contribution to the multiplicity in these two intervals is provided by
independent contributions from left- and right-moving wounded nucleons. It is
similar to the assumption of independent hadronization of strings in the dual
parton model \cite{DPM}. The picture of independent left- and right-moving
sources of particles is the main assumption of the wounded nucleon model.
There are many phenomenological and experimental evidences supporting this
idea \cite{ff-bc,ff-Ryb,bz-ff}.

It is convenient to construct the generating function%
\begin{equation}
H\left(  z_{B},z_{F}\right)  =\sum\limits_{n_{B},n_{F}}P\left(  n_{B}%
,n_{F}\right)  z_{B}^{n_{B}}z_{F}^{n_{F}}, \label{H1}%
\end{equation}
where $P\left(  n_{B},n_{F}\right)  $ is the probability to find $n_{B}$
particles in $B$ and $n_{F}$ in $F$. In general we may have many sources of
particles (wounded nucleons), thus $P\left(  n_{B},n_{F}\right)  $ can be
expressed as%
\begin{equation}
P\left(  n_{B},n_{F}\right)  =\sum_{w_{L},w_{R}}W\left(  w_{L},w_{R}\right)
P\left(  n_{B},n_{F};w_{L},w_{R}\right)  ,
\end{equation}
where $W\left(  w_{L},w_{R}\right)  $ is the probability distribution of the
numbers of wounded nucleons moving left $w_{L}$ and right $w_{R}$,
respectively. $P\left(  n_{B},n_{F};w_{L},w_{R}\right)  $ is the probability
to find $n_{B}$ particles in $B$ and $n_{F}$ in $F$ under condition of $w_{L}$
and $w_{R}$ wounded nucleons in left- and right-moving nucleus, respectively.

As derived in the Appendix, the generating function (\ref{H1}) reads
\begin{align}
H\left(  z_{B},z_{F}\right)   &  =\sum\limits_{w_{L},w_{R}}W\left(
w_{L},w_{R}\right)  \left\{  1+\frac{\bar{n}}{k}\left[  p_{LB}\left(
1-z_{B}\right)  +p_{LF}\left(  1-z_{F}\right)  \right]  \right\}  ^{-kw_{L}%
/2}\times\nonumber\\
&  \times\left\{  1+\frac{\bar{n}}{k}\left[  p_{RB}\left(  1-z_{B}\right)
+p_{RF}\left(  1-z_{F}\right)  \right]  \right\}  ^{-kw_{R}/2}, \label{H2}%
\end{align}
where $p_{RF}$ denotes the probability that a particle originating from the
right-moving wounded nucleon goes to $F$ interval, under the condition that
this particle was found either in $B$ or $F$ (and analogous for $p_{RB}%
,p_{LB}$ and $p_{LF}$). These probabilities satisfy natural conditions
\begin{equation}
p_{LB}+p_{LF}=1,\quad p_{RB}+p_{RF}=1. \label{p_suma}%
\end{equation}
These numbers can be easily calculated. Indeed, they depend only on positions
and sizes of $B$ and $F$ as well as the shape of the wounded nucleon
fragmentation function $\rho\left(  \eta\right)  $. For instance, $p_{RF}$ has
the form%
\begin{equation}
p_{RF}=\frac{\int_{F}\rho\left(  \eta\right)  d\eta}{\int_{B+F}\rho\left(
\eta\right)  d\eta}. \label{p_def}%
\end{equation}
The parameters $\bar{n}$ and $k$ come from the NB distribution fit (\ref{nbd})
to the $pp$ multiplicity distribution data in the combined interval $B+F$.
These parameters are well known for various energies and different
pseudorapidity intervals \cite{UA5-200-n-k}. Moreover
\begin{equation}
\bar{n}=2\int_{B+F}\rho\left(  \eta\right)  d\eta. \label{mn}%
\end{equation}
It is worth to notice that formula (\ref{H2}) contains all information about
the multiplicities in $B$ and $F$, as well as their dependence on the number
of wounded nucleons.

Using definitions (\ref{b_def}) and (\ref{H1}) we obtain [$b\equiv U/D$]:%
\begin{align}
U  &  =\left[  \frac{\partial^{2}H}{\partial z_{B}\partial z_{F}%
}-\frac{\partial H}{\partial z_{B}}\frac{\partial H}{\partial z_{F}}\right]
_{z_{B},z_{F}=1},\nonumber\\
D  &  =\left[  \frac{\partial^{2}H}{\partial z_{F}^{2}}+\frac{\partial
H}{\partial z_{F}}-\left(  \frac{\partial H}{\partial z_{F}}\right)
^{2}\right]  _{z_{B},z_{F}=1}.
\end{align}
Performing appropriate differentiations we obtain
\begin{align}
\frac{4U}{\bar{n}^{2}}  &  =p_{LB}p_{LF}\left[  \left\langle w_{L}%
^{2}\right\rangle -\left\langle w_{L}\right\rangle ^{2}+\frac{2\left\langle
w_{L}\right\rangle }{k}\right]  +p_{RB}p_{RF}\left[  \left\langle w_{R}%
^{2}\right\rangle -\left\langle w_{R}\right\rangle ^{2}+\frac{2\left\langle
w_{R}\right\rangle }{k}\right]  +\nonumber\\
&  +\left(  p_{LB}p_{RF}+p_{LF}p_{RB}\right)  \left[  \left\langle w_{L}%
w_{R}\right\rangle -\left\langle w_{L}\right\rangle \left\langle
w_{R}\right\rangle \right]  , \label{U}%
\end{align}
and
\begin{align}
\frac{4D}{\bar{n}^{2}}  &  =p_{LF}^{2}\left[  \left\langle w_{L}%
^{2}\right\rangle -\left\langle w_{L}\right\rangle ^{2}+\frac{2\left\langle
w_{L}\right\rangle }{k}\right]  +p_{RF}^{2}\left[  \left\langle w_{R}%
^{2}\right\rangle -\left\langle w_{R}\right\rangle ^{2}+\frac{2\left\langle
w_{R}\right\rangle }{k}\right]  +\nonumber\\
&  +2p_{LF}p_{RF}\left[  \left\langle w_{L}w_{R}\right\rangle -\left\langle
w_{L}\right\rangle \left\langle w_{R}\right\rangle \right]  +2\frac{p_{LF}%
\left\langle w_{L}\right\rangle +p_{RF}\left\langle w_{R}\right\rangle }%
{\bar{n}}. \label{D}%
\end{align}
In the above expressions $\left\langle ...\right\rangle $ represents the
average with respect to $W\left(  w_{L},w_{R}\right)  $. For instance
$\left\langle w_{L}\right\rangle $ is the average number of wounded nucleons
in the left-moving nucleus.

\section{Fully symmetric case}

The result presented in the previous section is valid for any colliding
nuclei. In case of symmetric collisions we of course have $\left\langle
w_{L}\right\rangle =\left\langle w_{R}\right\rangle $ and $\left\langle
w_{L}^{2}\right\rangle =\left\langle w_{R}^{2}\right\rangle $. Moreover,
studying correlations in symmetric (around $\eta=0$) intervals, i.e.,
$p_{LB}=p_{RF}\equiv p$ and $p_{LF}=p_{RB}=1-p$, we obtain%
\begin{equation}
b=\frac{C_{1}\left[  \left\langle w_{R}^{2}\right\rangle -\left\langle
w_{R}\right\rangle ^{2}+2\left\langle w_{R}\right\rangle /k\right]
+C_{2}\left[  \left\langle w_{L}w_{R}\right\rangle -\left\langle
w_{R}\right\rangle ^{2}\right]  }{C_{2}\left[  \left\langle w_{R}%
^{2}\right\rangle -\left\langle w_{R}\right\rangle ^{2}+2\left\langle
w_{R}\right\rangle /k\right]  +C_{1}\left[  \left\langle w_{L}w_{R}%
\right\rangle -\left\langle w_{R}\right\rangle ^{2}\right]  +2\left\langle
w_{R}\right\rangle /\bar{n}}, \label{b_sym}%
\end{equation}
where%
\begin{equation}
C_{1}=2p(1-p),\quad C_{2}=1-C_{1}. \label{C1C2}%
\end{equation}
It is worth to notice that formula (\ref{b_sym}) simplifies for two cases.

(i) Midrapidity. Considering two narrow rapidity intervals $B$ and $F$ around
$\eta=0$ we have $p\approx0.5$. It leads to a particularly simple expression%
\begin{equation}
b=1-\left[  1+\frac{\bar{n}}{4}\left(  \frac{2}{k}+\frac{\left\langle
w^{2}\right\rangle -\left\langle w\right\rangle ^{2}}{\left\langle
w\right\rangle }\right)  \right]  ^{-1}, \label{b_mid}%
\end{equation}
where $w=w_{L}+w_{R}$ is the number of wounded nucleons in both colliding
nuclei. This formula allows to notice the growth of $b$ with increasing scaled
variance of the number of wounded nucleons $\left\langle \left[
w-\left\langle w\right\rangle \right]  ^{2}\right\rangle /\left\langle
w\right\rangle $.

(ii) Fragmentation region. Assuming that intervals $B$ and $F$ are separated
enough so that $F$ can be populated only by right-moving wounded nucleons and
$B$ only by the left-moving ones, that is $p=1$, we obtain%
\begin{equation}
b=\frac{\left\langle w_{L}w_{R}\right\rangle -\left\langle w_{R}\right\rangle
^{2}}{\left\langle w_{R}\right\rangle }\left[  \frac{2}{\bar{n}}+\frac{2}%
{k}+\frac{\left\langle w_{R}^{2}\right\rangle -\left\langle w_{R}\right\rangle
^{2}}{\left\langle w_{R}\right\rangle }\right]  ^{-1}. \label{b_frag}%
\end{equation}
In this case $b>0$ only due to the fluctuations of the number of wounded
nucleons, i.e., $b=0$ if $\left\langle w_{L}w_{R}\right\rangle =\left\langle
w_{R}\right\rangle ^{2}$.

This closes the theoretical discussion of the problem.

\section{Results}

Recently the STAR collaboration presented results \cite{STAR} on correlation
coefficient $b$ for $AuAu$ collisions at $\sqrt{s}=200$ GeV. The backward
$B=(-\frac{\Delta\eta}{2}-0.1,-\frac{\Delta\eta}{2}+0.1)$ and forward
$F=(\frac{\Delta\eta}{2}-0.1,\frac{\Delta\eta}{2}+0.1)$ intervals of width
$0.2$ each were located symmetrically around $\eta=0$ with the distance
$\Delta\eta$ between bin centres ranging from $0.2$ to $1.8$ with an interval
of $0.2$. The measurement was performed for different centrality classes
defined via the number of produced particles in the central rapidity
region.\footnote{For instance $0-10\%$ centrality class corresponds to events
with the number of produced particles (in the central region) larger then
$430$ \cite{PHD}.}

As argued in Ref. \cite{WW} different centrality selections (e.g., via impact
parameter, number of wounded nucleons, number of produced particles) give the
same average number of wounded nucleons $\left\langle w\right\rangle $.
However, as was shown in Ref. \cite{Torr}, they lead to rather different
$\Omega\equiv\lbrack\left\langle w^{2}\right\rangle -\left\langle
w\right\rangle ^{2}]/\left\langle w\right\rangle $, except the most central
collisions, where $\Omega$ weakly depends on the centrality class definition.
In consequence, direct comparison of our result (\ref{b_sym}) with the STAR
data can be performed only for the most central collisions. In case of
non-central collisions the comparison is not straightforward. Indeed, wounded
nucleon model does not describe correctly the multiplicities in $AuAu$
collisions, thus not allowing to impose experimental centrality class cuts on
the number of produced particles.

We performed our calculations with the centrality class definition via the
number of wounded nucleons $w=w_{L}+w_{R}$ in both colliding nuclei (obviously
the impact parameter fluctuations are also included). We performed Monte-Carlo
calculations \cite{MC} for five centrality class selections: $0-10\%$,
$10-20\%$, $20-30\%$, $30-40\%$ and $40-50\%$ what correspond to $w\geq275,$
$275>w\geq197,$ $197>w\geq139,$ $139>w\geq94$ and $94>w\geq60$, respectively.
The corresponding results for $\left\langle w_{R}\right\rangle $,
$\left\langle w_{R}^{2}\right\rangle $, $\left\langle w_{L}w_{R}\right\rangle
$ and $\Omega$ are presented in Table \ref{Tab1}. \begin{table}[h]
\begin{center}%
\begin{tabular}
[c]{|c|c|c|c|c|}\hline\hline
$\%$ & $\left\langle w_{R}\right\rangle $ & $\left\langle w_{R}^{2}%
\right\rangle $ & $\left\langle w_{L}w_{R}\right\rangle $ & $\Omega
$\\\hline\hline
0-10 & 163 & 26841 & 26793 & 3.04\\\hline
10-20 & 116.8 & 13803 & 13723 & 2.07\\\hline
20-30 & 83.1 & 7015 & 6938 & 1.71\\\hline
30-40 & 57.5 & 3378 & 3315 & 1.40\\\hline
40-50 & 37.83 & 1478 & 1432 & 1.26\\\hline
\end{tabular}
\end{center}
\caption{The MC results for $\left\langle w_{R}\right\rangle $, $\left\langle
w_{R}^{2}\right\rangle $, $\left\langle w_{L}w_{R}\right\rangle $ and
$\Omega=[\left\langle w^{2}\right\rangle -\left\langle w\right\rangle
^{2}]/\left\langle w\right\rangle $ corresponding to five $10\%$ centrality
classes defined via the number of wounded nucleons $w=w_{L}+w_{R}$ in both
colliding nuclei.}%
\label{Tab1}%
\end{table}In our MC calculations for the nuclear density profile we took the
standard Woods-Saxon approximation with the nuclear radius $R=6.38$ fm and the
skin depth $d=0.535$ fm \cite{R}. For the nucleon-nucleon interaction profile
we used the black disk approximation\footnote{We also performed calculations
for the Gaussian approximation. We observe the week dependence of our results
on the $pp$ interaction profile.}, i.e., the interaction takes place only if
the transverse distance between two colliding nucleons is smaller than
$\sqrt{\sigma/\pi},$ with the total inelastic $pp$ cross section $\sigma=42$ mb.

\begin{figure}[h]
\begin{center}
\includegraphics[scale=1.5]{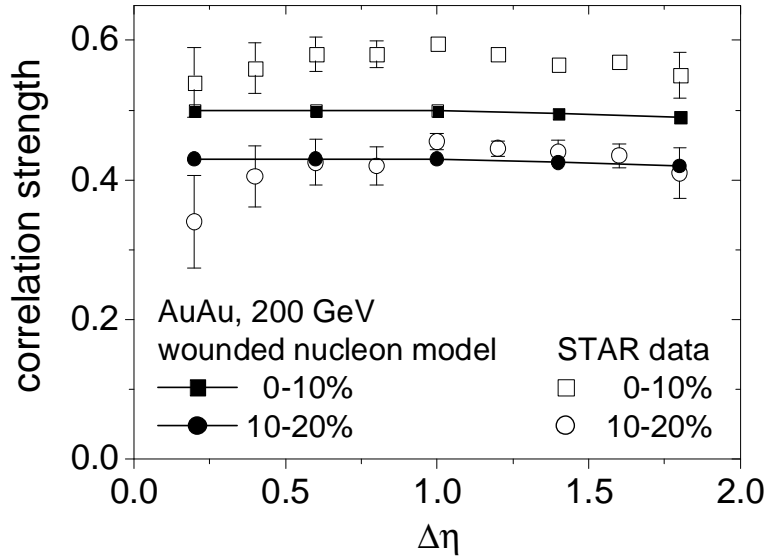}
\end{center}
\caption{The STAR data points compared with the results of the wounded nucleon
model for the correlation coefficient $b$ for two most central events vs. the
distance $\Delta\eta$ between bin centres. The width of each bin equals
$0.2$.}%
\label{fig_mid_1}%
\end{figure}

In Fig. \ref{fig_mid_1} the calculated correlation coefficient $b$
(\ref{b_sym}) for $0-10\%$ and $10-20\%$ vs. the distance $\Delta\eta$ between
bin centres is compared with the STAR data \cite{PHD}. Taking Eq.
(\ref{p_def}) into account we obtain $p=0.51,$ $0.52,$ $0.55,$ $0.56,$ $0.58$
for $\Delta\eta=0.2,$ $0.6,$ $1.0,$ $1.4,$ $1.8,$ respectively. NB
distribution fits to $pp$ multiplicity data in the midrapidity region give
approximately constant $\bar{n}=0.96$ (central plateau) and $k=1.8$
\cite{UA5-200-n-k}.\footnote{From Ref. \cite{UA5-200-n-k} it may be concluded
that $k$ is slightly increasing to $k\approx2$ for $\Delta\eta=1.8$. This
effect, however, practically does not influence numerical values of $b$.} It
is interesting to note that for the $0-10\%$ most central events, where direct
comparison with the data is possible, the wounded nucleon model can explain
more than $85\%$ of the effect.

\begin{figure}[h]
\begin{center}
\includegraphics[scale=1.5]{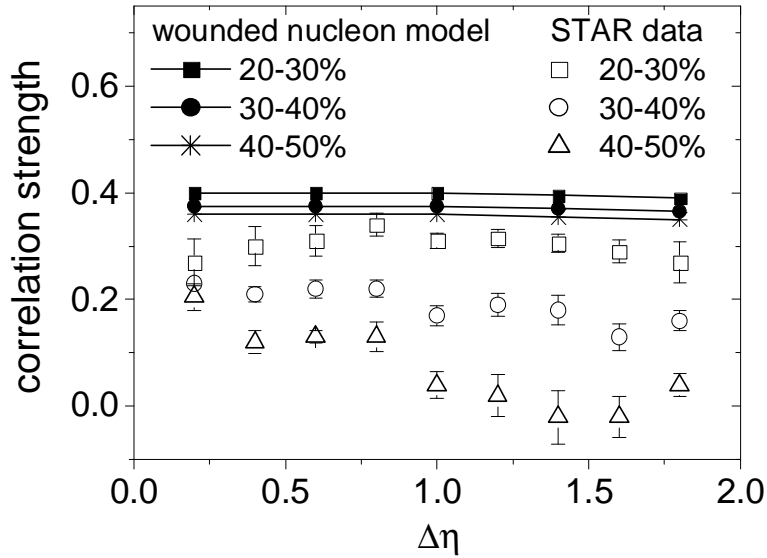}
\end{center}
\caption{The STAR data and the results of the wounded nucleon model for the
correlation coefficient $b$ for non-central events vs. the distance
$\Delta\eta$ between bin centres. As explained in the text in this case direct
comparison between data and model cannot be performed.}%
\label{fig_mid_2}%
\end{figure}

In Fig. \ref{fig_mid_2} the correlation coefficient $b$ (\ref{b_sym}) for
$20-30\%$, $30-40\%$ and $40-50\%$ centrality events vs. the distance
$\Delta\eta$ between bin centres is shown. The wounded nucleon model predicts
larger values of $b$ than observed, however, as explained at the beginning of
this section in case of non-central collisions the direct comparison with the
data cannot be performed [unknown precise value of $[\left\langle
w^{2}\right\rangle -\left\langle w\right\rangle ^{2}]/\left\langle
w\right\rangle $, see Eq. (\ref{b_mid})]. The main experimental finding,
however, that the correlation coefficient is approximately flat\footnote{This
is not true for the $40-50\%$ centrality events. We will come back to this
point in the next section.} in the midrapidity region as a function of
$\Delta\eta$ is very well reproduced in the wounded nucleon model. This
feature of the model can be easily understood from Eq. (\ref{b_mid}). Indeed,
$\bar{n}$ and $k$ are approximately constant (central plateau) in the
midrapidity region and the value of $p$ is close to $0.5$, which is a
consequence of the longitudinal structure of the wounded nucleon fragmentation
function, shown in Fig. \ref{fig_n}. 

\begin{figure}[h]
\begin{center}
\includegraphics[scale=1.5]{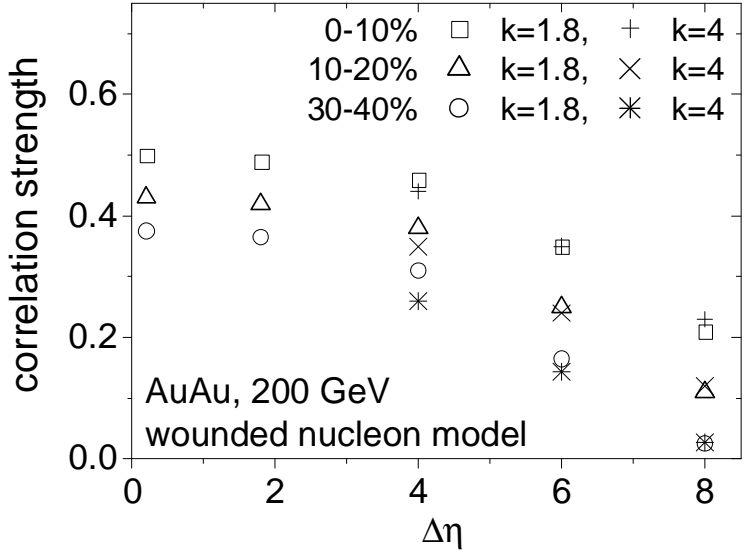}
\end{center}
\caption{Wounded nucleon model prediction for the correlation coefficient $b$
in the broad range of the distance $\Delta\eta$ between bin centres. The width
of each bin equals $0.2$.}%
\label{fig_full}%
\end{figure}

In Fig. \ref{fig_full} the correlation coefficient $b$ (\ref{b_sym}) for
$0-10\%$, $10-20\%$ and $30-40\%$ centrality events vs. the distance
$\Delta\eta$ in the broader range is shown. Taking Eqs. (\ref{p_def}) and
(\ref{mn}) we obtain $(p,$ $\bar{n})=(0.7,$ $0.97),$ $(0.85,$ $0.79),$ $(1,$
$0.53)$ for $\Delta\eta=4,$ $6,$ $8,$ respectively. The values of parameter
$k$ are not known precisely within these intervals, however, as can be
concluded from \cite{UA5-200-n-k} they should not be larger then $k=4$. As
shown in Fig. \ref{fig_full}, where the results are presented for $k=1.8$ and
$k=4$ for $\Delta\eta\geq4$, this uncertainty practically does not influence
our final results. The reduction of the correlation coefficient $b$ at
$\Delta\eta=8$ (at this point $b\approx0$ for peripheral collisions) is fully
determined by the suppression of particle production from a wounded nucleon to
the backward hemisphere\footnote{For instance, assuming that the contribution
from a wounded nucleon is symmetric around $\eta=0$ (i.e. $p=0.5$ at any
$\Delta\eta$) we would obtain $b\approx0.2$ at $\Delta\eta=8$ for $30-40\%$
centrality events.}, see Fig. \ref{fig_n}.

\section{Comments}

Following comments are in order.

(a) It is well-know that the wounded nucleon model significantly
underestimates the multiplicities in $AuAu$ collisions \cite{dataAuAu}.
Contrary to the model assumption multiplicity from a wounded nucleon depends
on the number of collisions it underwent. In order to take this effect into
account we multiplied $\bar{n}$ and $k$ by the ratio $\gamma$ of the measured
multiplicity in $AuAu$ collisions \cite{dataAuAu} to the prediction of the
wounded nucleon model $\bar{n}w/2$ \cite{WNM}. For the $0-10\%$ most central
collisions we approximately obtain $\gamma\approx1.6$ and consequently
$b\approx0.59$, which is in very good agreement with the measured value, see
Fig. \ref{fig_mid_1}.

(b) As seen in Figs. \ref{fig_mid_1}, \ref{fig_mid_2} and from Eq.
(\ref{b_mid}) the correlation coefficient calculated in the wounded nucleon
model is always flat in the midrapidity region in contrast to the $30-40\%$,
$40-50\%$ peripheral $AuAu$ or $pp$ collisions. In the present paper we
suggest that the fluctuation of the number of wounded nucleons may be
responsible for the large value of the forward-backward correlation
coefficient in central $AuAu$ collisions. For peripheral collisions, however,
this source of correlations is becoming less important (the value of
$[\left\langle w^{2}\right\rangle -\left\langle w\right\rangle ^{2}%
]/\left\langle w\right\rangle $ decreases) and obviously the mechanism
responsible for correlations in elementary $pp$ collisions play a major role.
In our approach we cannot describe the precise shape of the correlation
coefficient in $pp$ collisions\footnote{Neglecting fluctuations in Eq.
(\ref{b_mid}) we obtain in the midrapidity region the constant value of
$b\approx0.2$.} as a function of $\Delta\eta$, thus at the same time our
approach is not applicable to the peripheral $AuAu$ collisions.

(c) Encouraged by the success of our approach we also provide the prediction
for the correlation coefficient $b$ for the most central $PbPb$ collisions at
the LHC energy $\sqrt{s}=5500$ GeV. Performing appropriate MC
calculations\footnote{Here $R=6.62$ fm, $d=0.546$ fm and $\sigma=67$ mb
\cite{R}.} (described in the previous section) we obtained for the $0-10\%$
centrality events the following value of $[\left\langle w^{2}\right\rangle
-\left\langle w\right\rangle ^{2}]/\left\langle w\right\rangle =2.55$. Once
the parameters $\bar{n}$ and $k$ are measured at a given $B+F$ interval in
$pp$ collisions, the predictions for the correlation coefficient $b$ in the
midrapidity region can be easily obtained from Eq. (\ref{b_mid}). As an
example we present the result for $b$ in the midrapidity region with the
forward and backward intervals identical to those at the STAR measurement. The
needed parameters $\bar{n}\approx1.7$ \cite{Busza-lhc} and $k\approx1.5$
\cite{Dias-k,UA5-200-n-k} are taken from extrapolations to the LHC energy. In
consequence we obtain $b\approx0.6$ or $b\approx0.7$ if the correction to the
wounded nucleon model, discussed at the beginning of this section, is taken
into account. Our value is close to the prediction reported in Ref.
\cite{BDP-lhc}. However, in our approach the large value of $b$ is only due to
the fluctuation in the number of wounded nucleons, which if neglected (it
corresponds to $pp$ collisions) we obtain $b\approx0.35$ in contrast to the
value reported in Ref. \cite{BDP-lhc}.

(d) The wounded nucleon pseudorapidity fragmentation function, shown in Fig.
\ref{fig_n}, extends far beyond its own hemisphere. As discussed in the
previous section this feature is partially responsible for the approximately
constant value of $b$ in the midrapidity region. It is interesting to note
that similar longitudinal structure is present in the dual parton model (DPM)
\cite{DPM}, where the long longitudinally extended strings are stretched
between quarks and diquarks of the projectile and target, respectively. In
general, models that can explain the long-range forward-backward correlations
are models that introduce long extended objects in rapidity
\cite{DPM,CGC,BPV,BDP-lhc,GU,FK}. Moreover, in DPM the growth of the
correlation coefficient is due to the fluctuations in the number of elementary
inelastic collisions, which is similar to the fluctuations of the number of
wounded nucleons present in our approach. Therefore it is not surprising that
the two models lead to similar qualitative results \cite{STAR}. However, the
wounded nucleon model is in better agreement with data.

(e) Similar longitudinal structure provides the QCD inspired color glass
condensate model (CGC) \cite{CGC-old}, which includes many features of DPM. In
this approach \cite{CGC} the long extended color flux tubes and the
fluctuations of the number of gluons allow to understand the main features of
the STAR data. Moreover, it was shown recently \cite{cgc-ridge} that the soft
ridge structure observed at RHIC \cite{ridge} can be naturally understood in
the CGC/glasma motivated phenomenology, which is rather difficult to obtain in
the framework of the wounded nucleon model. This problem is currently under
our investigation.

(f) It would be interesting to perform similar calculation of the correlation
coefficient $b$ in the framework of the wounded quark-diquark model
\cite{ff-bb,wqdm}, which proved to be quite successful in description of
particle production in $pp$, $dAu$, $CuCu$ and $AuAu$ collisions. In this
model the number of produced particles is proportional to the number of
wounded quarks and diquarks, which are assumed to be the constituents of each
nucleon. Here the growth of the correlation coefficient is due to the
fluctuations of the number of wounded quarks and diquarks at a given
centrality bin.

(g) In the present approach we implicitly assume that particles are produced
directly from wounded nucleons. It would be interesting to check an effect of
intermediate resonances (clusters) production. We expect this effect to
influence the forward-backward multiplicity correlations in the midrapidity
region for peripheral $AuAu$ and $pp$ collisions.

\section{Conclusions}

Our conclusions can be formulated as follows.

(i) We have studied the forward-backward multiplicity correlations in the
framework of the wounded nucleon model \cite{WNM}. In this model particles are
produced independently from the left- and right-moving nucleons that
interacted in inelastic way at least once. An analytical expression for the
correlation coefficient (strength) for the collision of two arbitrary nuclei
and at any forward $F$ and backward $B$ intervals was derived.

(ii) The main ingredients of our approach are: recently obtained long extended
in pseudorapidity wounded nucleon fragmentation function \cite{ff-bc,ff-bb}
and the multiplicity distributions measured in proton-proton collisions
described by a negative binomial distribution.

(iii) In the midrapidity region correlation coefficient can be written in a
particularly simple form (\ref{b_mid}). This expression allows to explain the
growth of the correlation coefficient with increasing scaled variance
$\left\langle \left[  w-\left\langle w\right\rangle \right]  ^{2}\right\rangle
/\left\langle w\right\rangle $ of the number of wounded nucleons $w$ in both
colliding nuclei.

(iv) We have performed explicit calculations for $AuAu$ collisions at
$\sqrt{s}=200$ GeV. The backward/forward intervals were chosen according to
the recent STAR measurement. Growth of the correlation coefficient with
centrality as well as almost no pseudorapidity dependence in the midrapidity
region was observed. Our results are in good qualitative agreement with the
STAR data, although exact comparison can be performed only for the most
central collisions.

(v) Finally, predictions for the values of the correlation coefficient in the
broad range of pseudorapidity were presented.

\bigskip

\textbf{Acknowledgements}

We would like to thank Andrzej Bia\l as for suggesting this investigation and
useful discussions. Discussions with Piotr Bo\.{z}ek on the Monte Carlo
methods are highly appreciated. We also thank Wojciech Florkowski for
discussions on Ref. \cite{WW}. This investigation was supported in part by the
Polish Ministry of Science and Higher Education, grant No. N202 034 32/0918.

\appendix                                  

\section{Appendix: Generating function}

Let $P_{L}(n_{LB},n_{LF})$ be the probability that a left-moving wounded
nucleon contributes $n_{LB}$ particles into $B$ and $n_{LF}$ particles into
$F$ interval [and analogous distribution $P_{R}(n_{RB},n_{RF})$ for a
right-moving source]. The probability to find $n_{B}=$ $n_{LB}+n_{RB}$
particles in $B$ and $n_{F}=$ $n_{LF}+n_{RF}$ particles in $F$ in case of one
left- and one right-moving wounded nucleon is given by
\begin{equation}
P(n_{B},n_{F})=\sum_{\substack{n_{LB},n_{LF}\\n_{RB},n_{RF}}}P_{L}%
(n_{LB},n_{LF})P_{R}(n_{RB},n_{RF})\delta_{n_{LB}+n_{RB}}^{n_{B}}%
\delta_{n_{LF}+n_{RF}}^{n_{F}},
\end{equation}
and the generating function ($1,1$ means one left- and one right-moving
wounded nucleon)
\begin{equation}
H\left(  z_{B},z_{F};1,1\right)  =\sum\limits_{n_{B},n_{F}}P\left(
n_{B},n_{F}\right)  z_{B}^{n_{B}}z_{F}^{n_{F}}=H_{L}\left(  z_{B}%
,z_{F}\right)  H_{R}\left(  z_{B},z_{F}\right)  ,
\end{equation}
with
\begin{align}
H_{L}\left(  z_{B},z_{F}\right)   &  =\sum_{n_{LB},n_{LF}}P_{L}(n_{LB}%
,n_{LF})z_{B}^{n_{LB}}z_{F}^{n_{LF}},\nonumber\\
H_{R}\left(  z_{B},z_{F}\right)   &  =\sum_{n_{RB},n_{RF}}P_{R}(n_{RB}%
,n_{RF})z_{B}^{n_{RB}}z_{F}^{n_{RF}}.
\end{align}
It is easy to check that in case of $w_{L}$ left-moving and $w_{R}$
right-moving wounded nucleons we obtain
\begin{equation}
H\left(  z_{B},z_{F};w_{L},w_{R}\right)  =\left[  H_{L}\left(  z_{B}%
,z_{F}\right)  \right]  ^{w_{L}}\left[  H_{R}\left(  z_{B},z_{F}\right)
\right]  ^{w_{R}}.
\end{equation}

Suppose that $P_{1}(n)$ is the multiplicity distribution from a single wounded
nucleon in the combined interval $B+F$. Then
\begin{equation}
P_{L}(n_{LB},n_{LF})=P_{1}(n=n_{LB}+n_{LF})\frac{\left(  n_{LB}+n_{LF}\right)
!}{n_{LB}!n_{LF}!}\left(  p_{LB}\right)  ^{n_{LB}}\left(  p_{LF}\right)
^{n_{LF}},
\end{equation}
where $p_{LB}$ and $p_{LF}$ are defined in section $2$. In consequence%
\begin{equation}
H_{L}\left(  z_{B},z_{F}\right)  =\sum\nolimits_{n}P_{1}(n)\left[  p_{LB}%
z_{B}+p_{LF}z_{F}\right]  ^{n}.
\end{equation}
Performing analogous calculations for the right moving part
\begin{equation}
H_{R}\left(  z_{B},z_{F}\right)  =\sum\nolimits_{n}P_{1}(n)\left[  p_{RB}%
z_{B}+p_{RF}z_{F}\right]  ^{n}.
\end{equation}

Assuming that the multiplicity distribution measured in proton-proton
collision is described by a NB distribution with $\bar{n}$ and $k$ (in the
combined interval $B+F$), it is easy to show that $P_{1}(n)$ (from a wounded
nucleon) is given by a NB distribution with $\bar{n}/2$ and $k/2$ \cite{bm}.
Then using
\begin{equation}
\sum\nolimits_{n}P_{1}(n)\xi^{n}=\left(  1+\frac{\bar{n}(1-\xi)}{k}\right)
^{-k/2},
\end{equation}
we obtain
\begin{align}
H\left(  z_{B},z_{F};w_{L},w_{R}\right)   &  =\left\{  1+\frac{\bar{n}}%
{k}\left[  1-p_{LB}z_{B}-p_{LF}z_{F}\right]  \right\}  ^{-kw_{L}/2}%
\times\nonumber\\
&  \times\left\{  1+\frac{\bar{n}}{k}\left[  1-p_{RB}z_{B}-p_{RF}z_{F}\right]
\right\}  ^{-kw_{R}/2}.
\end{align}
Summing over $W(w_{L},w_{R})$, i.e., the probability distribution of the
number of wounded nucleons, and taking (\ref{p_suma}) into account we finally
obtain (\ref{H2}).

\end{document}